\documentclass[10pt,twocolumn,
    aps,prl,amsfonts,amssymb,
    superscriptaddress,floatfix,
    longbibliography]{revtex4-2}
\usepackage{graphicx}
\usepackage{wrapfig}
\usepackage{xcolor}
\usepackage{amsmath}
\usepackage{orcidlink}
\usepackage{mathrsfs}   
\usepackage{cancel}
\usepackage{physics}
\graphicspath{ {./figures/} }
\usepackage{hyperref}
\hypersetup{colorlinks=true,citecolor=blue,linkcolor=blue,urlcolor=blue}

\newcommand{\calH}{\mathcal{H}}

\begin{document}

\title{Flexible ground-state search algorithm on hybrid Clifford-tensor network manifolds}

\author{Hari Timsina~\orcidlink{0009-0003-1044-9856}}
\email{htimsina@sissa.it}
\affiliation{International School for Advanced Studies (\href{https://ror.org/004fze387}{SISSA}), Via Bonomea, 265, 34136 Trieste, Italy}

\author{Antonio Francesco Mello~\orcidlink{0009-0002-4401-1126}}
\thanks{Now at JPMorganChase.}
\affiliation{Center for Computational Quantum Physics, \href{https://ror.org/00sekdz59}{Flatiron Institute}, New York, NY 10010, USA}
\affiliation{International School for Advanced Studies (\href{https://ror.org/004fze387}{SISSA}), Via Bonomea, 265, 34136 Trieste, Italy}

\author{Meghadeepa Adhikary~\orcidlink{0000-0003-0427-2989}}
\affiliation{International School for Advanced Studies (\href{https://ror.org/004fze387}{SISSA}), Via Bonomea, 265, 34136 Trieste, Italy}

\author{Marcello Dalmonte~\orcidlink{0000-0001-5338-4181}}
\affiliation{The Abdus Salam International Centre for Theoretical Physics (\href{https://ror.org/009gyvm78}{ICTP}), Strada Costiera 11, 34151 Trieste, Italy}
\affiliation{Dipartimento di Fisica e Astronomia, \href{https://ror.org/01111rn36}{Universit\`a di Bologna}, and \href{https://ror.org/04j0x0h93}{INFN}, Sezione di Bologna, via Irnerio 46, I-40126 Bologna, Italy}

\author{Mario Collura~\orcidlink{0000-0003-2615-8140}}
\affiliation{International School for Advanced Studies (\href{https://ror.org/004fze387}{SISSA}), Via Bonomea, 265, 34136 Trieste, Italy}
\affiliation{\href{https://ror.org/05j3snm48}{INFN}, Sezione di Trieste, Via Valerio 2, 34127 Trieste, Italy}

\begin{abstract}
    Tensor network algorithms face challenges in highly entangled regimes, such as systems with periodic boundary conditions (PBC), long-range interactions, or higher dimensions, where the required bond dimension quickly renders computations infeasible. To overcome this limitation, we introduce a novel approach that integrates the stabilizer formalism directly into the traditional density-matrix renormalization group (DMRG) algorithm. While previous methods have used Clifford unitaries to minimize ground-state entanglement, fundamentally limited to distilling that away utilizing local two-qubit operations, our framework optimizes Clifford transformations to minimize the state's energy. We benchmark our method against standard DMRG at fixed bond dimensions for the Ising model on a ring (including long-range interactions) and on a torus. We observe systematic gains in models with local interactions - as drastic as a factor of 40 decrease in bond dimension in two dimensions. Our algorithm is readily implementable in standard DMRG codes and provides a practical way to substantially reduce algorithm costs, paving the way for large-scale simulation at volumes beyond what is presently achievable.
\end{abstract}
\maketitle

\textit{Introduction}---The solution of quantum many-body systems is, in general, an extremely challenging task due to the exponential growth of the size of Hilbert space with the number of particles~\cite{Feynman1982}. 
Exact solutions are limited to special, highly structured models and in smaller systems~\cite{Bethe1931, lieb_two_1961, baxter1985exactly}, whereas larger systems often call for computational  approaches~\cite{metropolis_equation_1953, RevModPhys.47.773, PhysRevLett.69.2863, carleo2017nqs}. The latter choice depends on the physical problem and the properties of interest. 
One of the most successful numerical approaches for low-dimensional strongly correlated systems is the Density Matrix Renormalization Group (DMRG)~\cite{PhysRevLett.69.2863, PhysRevB.48.10345, schollwock2005density, Schollw_ck_2011}, an iterative variational procedure that provides highly accurate approximations of the low-lying states (especially the ground state) by systematically optimizing the truncated Hilbert space according to the entanglement structure of the system. DMRG can be formulated within the tensor network (TN) framework, where one recasts the problem of a large many-body system into a network of smaller interconnected tensors that efficiently capture relevant quantum correlations and enable the study of systems that are otherwise computationally inaccessible. 
In one-dimensional TN, the ground state of quantum systems can be represented in the form of Matrix Product States (MPS), and operators are written as Matrix Product Operators (MPO)~\cite{PhysRevB.73.094423, Pirvu_2010, RevModPhys.93.045003, orus2014practical, crosswhite2008automata, frowis2010tensor, hubig2017mpo}. This decomposition introduces a controlled approximation through the dimensions of the auxiliary links connecting neighboring tensors, known as the bond dimensions of the network. 
The bond dimension determines how much entanglement can be captured across a given bond and therefore directly controls the accuracy of the TN representation.
\begin{figure*}[ht!]
    \centering
    \includegraphics[width=1.0\linewidth]{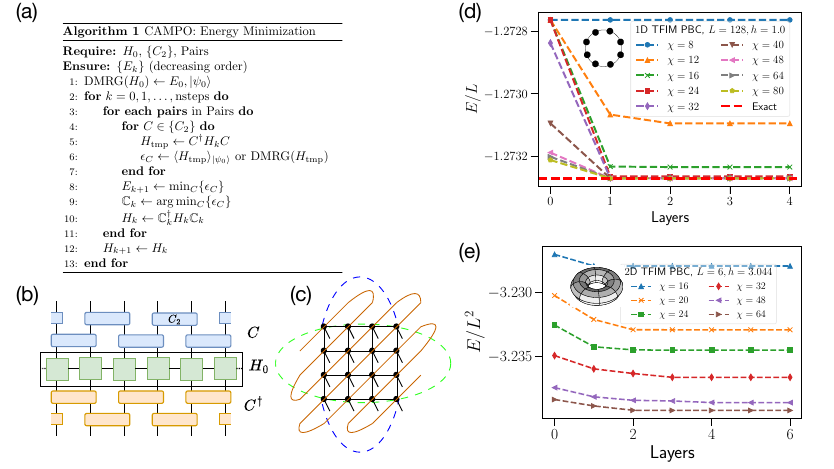}
    \caption{Summary of the present work. (a) The algorithm for the energy minimization with Clifford-augmented MPO. $\{C_2\}$ is the set of two-qubit Clifford operators, the number of which is 20 for the present purpose. The `Pairs' consists of the collection of site pairs where we apply the Clifford transformation in the Hamiltonian. The nsteps are chosen until the convergence of the successive energy, i.e., all Identity Clifford operator selections for all pairs in the update procedure. (b) Augmenting the Hamiltonian MPO (region inside the center rectangle) with Cliffords in a brickwall manner. The circuit efficiently performs the transformation $H_0\rightarrow C^\dagger H_0 C$. (c) Transformation of 2D geometry to the MPS with diagonal snaking. The two ovals indicate that periodic boundary conditions are employed in both directions. (d) Benchmarking the algorithms for the 1D TFIM with PBC for $L=128$ at the critical point, and the results are compared with the free-fermion (exact) result. (e) Energy minimization for the 2D TFIM at the critical point, where the results for the $6\times 6$ torus are shown for various bond dimensions.}
    \label{fig:summary}
\end{figure*}

TN methods have proven to be remarkably efficient for one-dimensional gapped systems, fundamentally because their entanglement entropy satisfies an area law~\cite{Hastings_2007, RevModPhys.82.277} $S\sim\partial A$ with $\partial A$ being the boundary separating a region $A$ from its complement $\bar{A}$. In this regime, the ground-state manifold can be precisely represented using relatively small bond dimensions--a paradigm that excels for locally interacting models with open boundary conditions~\cite{Verstraete01032008}.
However, determining the ground state and accurately computing its energy remains a nontrivial challenge in generic many-body systems. In particular, standard DMRG and related tensor network techniques may lose efficiency or accuracy in the presence of long-range interactions, periodic boundary conditions, or higher-dimensional geometries, where entanglement can grow substantially. In such cases, an accurate representation would require exponentially large bond dimensions, which is computationally infeasible.
This naturally raises the question of whether one can construct representations that yield improved ground-state approximations compared to conventional DMRG at the same bond dimension.

In this work, we address this question by combining DMRG with the stabilizer formalism and showing how the latter lets us target entanglement contributions of geometrical origin. While previous efforts have integrated Clifford unitaries into DMRG in an attempt to factor out the classical correlations in the ground state, we here adopt a different approach. Instead of building the Clifford-augmented matrix product state (CAMPS)~\cite{PhysRevLett.133.190402,lami2025quantumstate, mishmash2023hierarchical} approximated ground state by choosing Clifford circuits that lower its entanglement~\cite{PhysRevLett.133.190402, PhysRevB.111.085121}, we dress the DMRG algorithm with Clifford transformations that lower the energy of the optimized state~\cite{ravi2022cafqa, ryabinkin2018qcc, brown2025qcliff, brown2026hamiltonian, krumnow2016orbital, zgid2008orbital, haghshenas2022circuit} - that is, we work directly on the Clifford augmented matrix product operator (CAMPO) (see Fig.\ref{fig:summary}a-b). This allows us to retain full spatial flexibility of our augmentation: in particular, we can easily integrate long-range disentanglers (Fig.\ref{fig:summary}c), which were unavailable in previous approaches, and that allow us to systematically target entropic contributions due to boundary conditions at essentially no extra cost. 

We benchmark our method against standard DMRG at fixed bond dimensions on the 1D and 2D critical Ising models, and the 1D long-range Ising model with PBC (Fig.\ref{fig:summary}d-e). The improvement in the estimated energy is striking in the 1D models, where the proposed algorithm consistently outperforms plain DMRG for all system sizes, with larger margins observed at higher bond dimensions. A similar behavior is observed for the 2D case, with systematic improvements at criticality and in gapped phases. Overall, our findings underscore the efficacy of our algorithm in handling critical regimes within the CAMPS framework, while also paving the way for alternative optimization strategies of the ansatz.

\textit{CAMPS framework}---Let $I, X, Y, Z$ denote the identity and the three Pauli matrices, from which we can construct the $L$-qubit Pauli group $\mathcal{P}_L\equiv\{\pm 1,\pm i\}\times \{I, X, Y, Z\}^{\otimes L}$. Within the space of unitary operators, there exists a subset that maps Pauli operators to Pauli operators under conjugation: $\{U:UPU^\dagger\in \mathcal{P}_L, \forall P\in\mathcal{P}_L\}$. Such operators are known as normalizers of the Pauli group, and the collection of them forms the Clifford group. The Clifford group can be generated by three sets of operators (or gates), namely the Hadamard ($H$), the phase ($S$), and the controlled-NOT (\texttt{CNOT}) gates~\cite{nielsen2010quantum}. 
Although Clifford operations are capable of generating highly entangled quantum states, they nevertheless admit efficient classical simulation~\cite{gottesman1997stabilizercodesquantumerror, gottesman1998heisenbergrepresentationquantumcomputers, aaronson2004improved, Bravyi2019simulationofquantum}. Moreover, their unitary structure implies that the entanglement created by Clifford circuits can, in principle, be systematically disentangled using the same gate set. However, Clifford operations alone are not universal for quantum computation. To achieve universality, one must supplement them with at least one non-Clifford gate, most commonly the $T$ gate. This universality, however, comes at a computational cost characterized by the degree of nonstabilizerness (aka magic), which quantifies how far a quantum system is from the efficient classical simulation of the stabilizer manifold~\cite{bravyi2005universal, veitch2014resource, howard2017application, leone2022stabilizer, oliviero2022ising, haug2023quantifying, lami2023perfect, tarabunga2023manybody, tarabunga2024pauli, frau2024nonstabilizerness, tirrito2024flatness, mello2025retrieving, sinibaldi2025nonstabilizerness}.

The idea of evolving the MPS with suitably chosen Clifford operations has led to the CAMPS~\cite{mishmash2023hierarchical, PhysRevLett.133.190402, lami2025quantumstate}. 
In the CAMPS framework, Clifford transformations are adaptively optimized to minimize entanglement during the evolution, thereby reducing the required bond dimension and substantially extending the reach of tensor network simulations to larger system sizes~\cite{llima2024stabtn, mello2024stabmpo, mello2025ctdvp, mello2025loschmidt, qin2025ctdvp, qian2025finiteT, huang2025campsfermions, huang2026matchgates, fu2025cadmrg, yosprakob2025cagmps, li2026kitaev, kim2026haldane, chang2026disentanglers, masotllima2026limits}. 
A key advantage of the CAMPS procedure is that it leaves the nonstabilizerness of the system invariant, allowing one to reduce entanglement complexity without altering the intrinsic quantum resources associated with magic~\cite{frau2025stabilizer}.

\textit{Clifford updates for the MPO}---In the CAMPS framework, the expectation value of an operator ($O$) is given by $\bra{\psi_U}O\ket{\psi_U}$, where the state evolves in the Schr\"odinger picture accompanied by the Clifford unitary operator $U$ as $\ket{\psi_U}\equiv U\ket{\psi}$. Equivalently, one may adopt the Heisenberg picture, where the state remains fixed, and the operator evolves as $O_U\equiv U^\dagger OU$, thereby giving the expectation value $\bra{\psi}O_U\ket{\psi}$. 
Furthermore, if the operator is represented as an MPO, the CAMPS procedure can be reformulated by augmenting the Clifford circuit with the MPO representation, shifting the evolution from the quantum state to the operator, leaving the expectation value of the operator unchanged. This modified scheme enables computations to be performed within the manifold of an MPS with a fixed bond dimension.

In this regard, the computation of energy can be expressed as  
\begin{equation}
    E_U = \bra{\psi_U}\calH\ket{\psi_U} = \bra{\psi}\calH_U\ket{\psi},
    \label{eq:energy_U}
\end{equation}
where $\calH$ is the original Hamiltonian and $\calH_U=U^\dagger\calH U$ is the Clifford augmented Hamiltonian. 
The goal is to minimize the initial DMRG energy by iteratively applying Eq.~\eqref{eq:energy_U} and selecting the Clifford operators that yield the lowest energy.
This problem can also be reformulated in terms of the following variational technique. The variational parameter for the DMRG is the MPS $\ket{\psi}$, which is optimized through sweeps over the whole system such that we have the best state $\ket{\psi'}$ satisfying $\partial_{|\psi'\rangle^{*}} \langle\psi'|\calH|\psi'\rangle = 0$.  Updating the Hamiltonian and choosing the one with minimum energy is now equivalent to the variational problem 
\begin{equation} \label{eq:variationalC}
    \partial_{U} \langle \psi|\calH_U|\psi\rangle = 0,\; \text{ for }  U\in \text{Cliffords}.
\end{equation}

After the initial DMRG, the energy minimization can proceed in two ways: (i) by evaluating the expectation value of $\calH_U$ against the ground state of the initial DMRG, which is kept fixed throughout, or (ii) by re-running DMRG on $\calH_U$ after each Clifford update. Both target the same variational fixed point, but since the joint problem is non-convex, the two schedules generically follow different trajectories and need not converge to the same optimum. We find the second approach performs better and adopt it for the rest of the paper.

We restrict our study to spin--1/2 systems whose Hamiltonian can be expressed as a linear combination of Pauli operators, $H = \sum_{P\in\mathcal{P}_N} c_P P$.
The evolution of such a Hamiltonian under the Clifford unitaries can be efficiently tracked with the stabilizer formalism, and the number of terms in the sum remains fixed at each step of the evolution. Note that the Hermiticity of the original Hamiltonian is preserved during the Clifford updates.
Remarkably, while Clifford operations on an arbitrary number of qubits can be used for efficient updates, the number of such operations grows super-exponentially with system size. Therefore, we restrict ourselves to successive applications of two-qubit Clifford gates. Knowing that Clifford operations can effectively disentangle MPS, and that there are $20$ nonequivalent representations of two-qubit Cliffords, we use these same unitaries for the energy minimization algorithm. The energy minimization is performed after each application of a Clifford operation across the entire lattice. The procedure of energy minimization by updating the Hamiltonian with Cliffords is summarized as an algorithm in Fig.~\ref{fig:summary}(a), whereas the updating Hamiltonian itself is illustrated in Fig.~\ref{fig:summary}(b).
In contrast, other methods for operator evolution in the Heisenberg picture--such as Pauli propagation~\cite{rudolph2025pauli, begusic2024spd, angrisani2025classically} with arbitrary rotations-- generally lead to an exponential growth in the number of terms in the Hamiltonian at each step, necessitating the use of optimized truncation strategies to control the operator complexity.
The algorithm described above is benchmarked for different models as discussed below.

\textit{1D TFIM with PBC}---We first consider the integrable one-dimensional transverse field Ising model (1D TFIM) of $L$ sites with periodic boundary conditions
\begin{equation}
    \calH_{\rm 1D} = -J\sum_{i=1}^L X_i X_{i+1} - \sum_{i=1}^L hZ_i, \quad X_{L+1}=X_1.
    \label{eq:tfim1d}
\end{equation}
The Hamiltonian is translationally invariant, with nearest-neighbor interactions between adjacent spins along the chain.
The model of Eq.~\eqref{eq:tfim1d} exhibits a phase transition at zero temperature mediated by the transverse field~\cite{chakrabarti1996quantum}, with the quantum critical point $h/J=1.0$. The exact result for the ground-state energy from the free fermion theory with antiperiodic boundary conditions is given by~\cite{pfeuty1970onedimensional,mbeng2024quantum}:
\begin{equation}
E_{ex}=-\sum_{m=0}^{L-1} \sqrt{h^2+J^2-2hJ\cos\left(\tfrac{(2m+1)\pi}{L}\right)} 
\end{equation}
and this value is taken as the reference energy to compare with the results from the CAMPO algorithm. We set $J=1$ for convenience. 

The periodic geometry introduces long-range correlations for MPS representations, making it difficult for DMRG to converge to the ground-state energy, even for moderate bond dimensions~\cite{verstraete2004pbc, pippan2010pbc}. 
Applying DMRG to Eq.~\eqref{eq:tfim1d} yields an energy $E_0$ together with the corresponding state $\ket{\psi}$. We subsequently update the Hamiltonian through the application of two-qubit Clifford operations acting on pairs $(i,i+1), i=1,2,\cdots, L$ (including the boundary pair which is far apart in the MPS) in the standard brick-wall pattern presented in Fig.~\ref{fig:summary}(b).
For each pair, all possible Clifford operations are examined, and the Hamiltonian is updated using the Clifford transformation that minimizes the energy for that pair before proceeding to the next pair. A complete optimization over all pairs defines one layer of the algorithm, and the procedure is iterated over multiple layers until convergence of the energy for successive layers is achieved. 

The benchmarking is performed over different system sizes of 1D TFIM and for different fixed bond dimensions. The case of $L=128$ at the critical point for different values of initial bond dimensions is shown in Fig~\ref{fig:summary}(d). We found that, although the initial DMRG is far from providing a good approximation of the exact ground-state energy, the Clifford optimization procedure indeed allows states with $\chi >40$ to approach the free fermions result with much better accuracy. 
The improvement in the energy is found to depend on both the bond dimension, which is kept fixed throughout the simulation, and the specific pair of sites selected for the Hamiltonian update. 

Furthermore, the procedure reveals that the dominant contribution arises from updates involving the boundary terms. This observation indicates that boundary effects play a significant role in the convergence behavior of DMRG, highlighting their influence on the efficiency of the optimization process. 
As the depth of the Clifford circuit increases, the initially 2-body and 1-body local terms of the Hamiltonian ~\eqref{eq:tfim1d} are stretched into highly non-local Pauli strings spanning multiple sites across the lattice.
\begin{figure}[t!]
    \centering
    \includegraphics[width=1.0\linewidth]{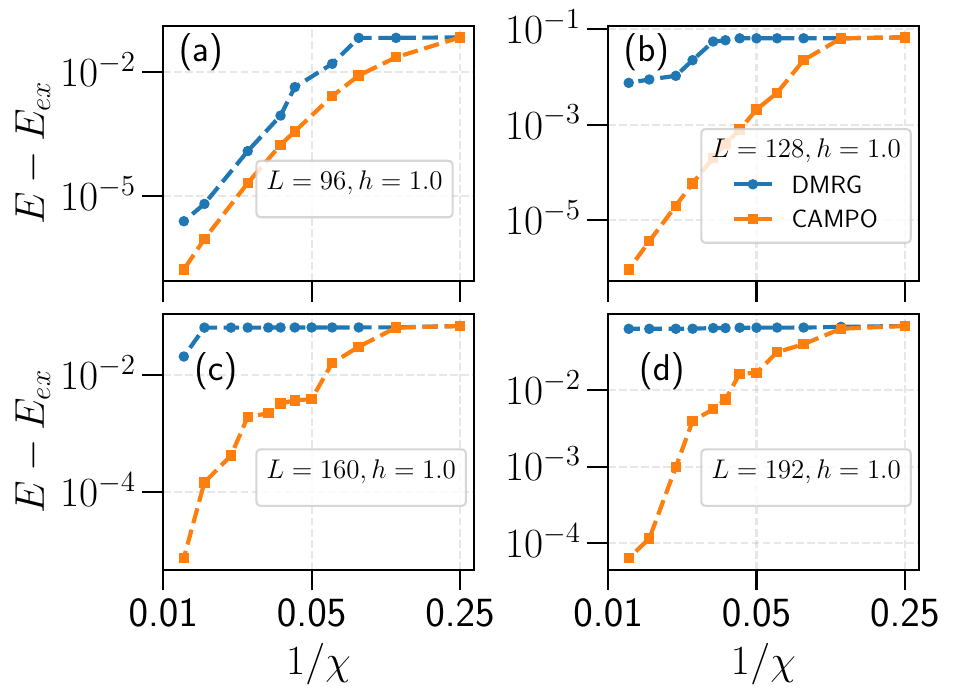}
    \caption{The scaling with bond dimension of the initial DMRG energy and the converged one with the Clifford optimization procedure for 1D TFIM PBC for different system sizes at the critical point $h=1.0$. Both results are subtracted from the free fermion result ($E_{ex}$) reported in the text. For fixed bond dimensions, the optimized results depend on the system sizes.}
    \label{fig:energyvs1byD}
\end{figure}

We also analyze convergence with respect to bond dimension by comparing the initial DMRG result with the final optimized energy. Results for different system sizes at the critical point are presented in Fig.~\ref{fig:energyvs1byD}, where both the DMRG and CAMPO curves represent the difference from the exact result with free fermions. 
The energy obtained from the DMRG for a particular bond dimension loses precision as we increase the system size; for example, with $\chi=80$, the energy is within $\sim 10^{-6}$ for $L=96$ (see Fig.~\ref{fig:energyvs1byD}(a)), whereas that becomes $\sim 10^{-1}$ for $L=192$ (Fig.~\ref{fig:energyvs1byD}(d)), the energy accuracy is $\sim 10^{-2}$ for intermediate system sizes as seen in Figs.~\ref{fig:energyvs1byD} (b) and (c) for $L=128$ and $160$ respectively.  Note that DMRG runs for different system sizes and bond dimensions are performed with the same parameters for the cutoff and number of sweeps. The degree of improvement for a fixed bond dimension is also found to depend on the size of the system, as seen from Fig.~\ref{fig:energyvs1byD}, with an accuracy gain of over $\sim 4$ orders of magnitude for the system size considered, as for the intermediate bond dimension. For comparison, the optimized energy with $\chi=80$ (leftmost point in each subfigure of Fig.~\ref{fig:energyvs1byD}) exceeds the single DMRG energy of the $\chi=1000$ value. It is also observed that the difference between the DMRG and CAMPO results increases with the bond dimension $\chi$. However, we expect this discrepancy to eventually decrease beyond a sufficiently large size-dependent $\chi$, as the DMRG results approach the exact value, leaving progressively less room for further improvement.

\textit{2D TFIM PBC}---The two-dimensional transverse field Ising model on a square lattice of size $L \times L$ is defined by the Hamiltonian,
\begin{equation}
    \calH_{\rm 2D} = -\sum_{i,j=1}^{L} \left(X_{i,j}X_{i+1,j} + X_{i,j}X_{i,j+1} \right) -h\sum_{i=1}^{L^2} Z_i.
    \label{eq:tfim2d}
\end{equation}

\begin{figure*}[t!]
    \centering
    \includegraphics[width=\textwidth]{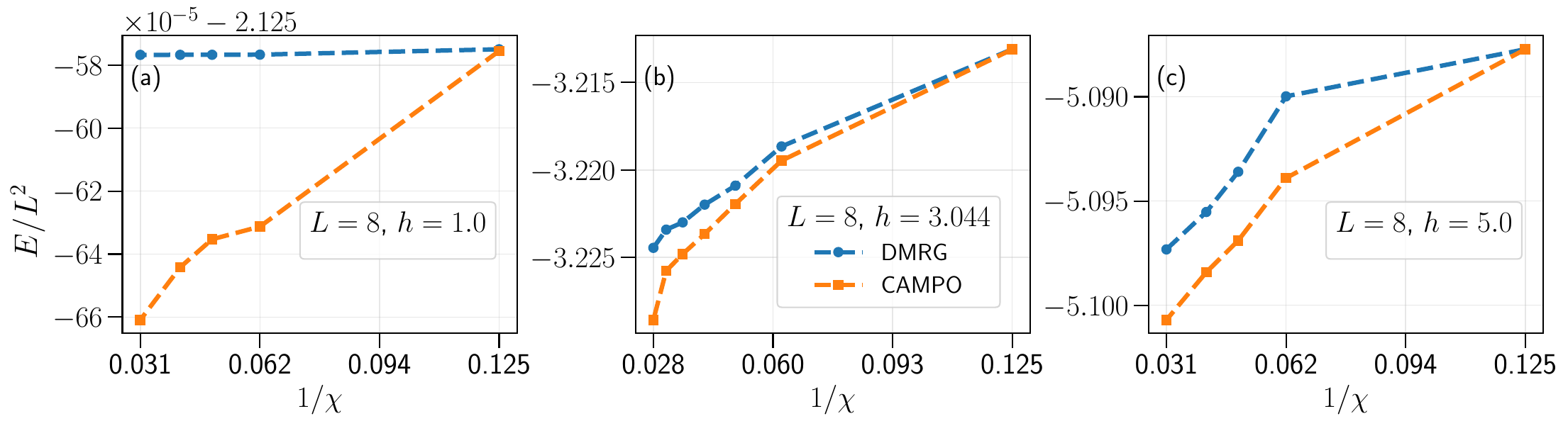}
    \caption{Scaling of the ground-state energy density with the inverse bond dimension $\chi$ for the transverse field Ising model on an $8 \times 8$ square lattice with torus geometry. Plain DMRG and Clifford-augmented MPO results are shown for different values of the transverse field \textbf{(a)} $h=1 < h_c$, \textbf{(b)} $ h=h_c$, and \textbf{(c)} $h=5>h_c$, where different order of improvement is observed for different field values. }
    \label{fig:energy_vs_BD_2d}
\end{figure*}
The model described by Eq.~\eqref{eq:tfim2d} is generally non-integrable, although numerical evidence from Monte Carlo simulations indicates a critical point at $h_c=3.04438(2)$~\cite{PhysRevE.66.066110, PhysRevB.57.8494, Hamer_2000, PhysRevB.106.054402, sandvik2003sse}. 
While tensor-network methods such as projected entangled pair states (PEPS) provide a natural framework for two-dimensional systems~\cite{Verstraete01032008, RevModPhys.93.045003, orus2014practical, krinitsin2025ttn, sadrzadeh2019ttn}, they incur prohibitively high computational costs as compared to the 1D model even for systems with the same number of spins.
We therefore adopt an MPO representation of the Hamiltonian and an MPS ansatz for the wave function by mapping the two-dimensional lattice to a one-dimensional chain via a snaking ordering~\cite{stoudenmire2012two}. The specific mapping by diagonal snaking used in this work is illustrated in Fig.~\ref{fig:summary}(c). 
In this representation, local nearest-neighbor couplings in 2D are transformed into long-range interactions in the effective 1D description. (Note that the area-law entanglement of the 2D lattice becomes a volume-law scaling in the 1D MPS mapping, making accurate representation with a finite small bond dimension challenging; as a result, DMRG energies can remain far from the exact values.)
Following the 1D procedure, we apply Clifford updates to nearest-neighbor pairs across the 2D lattice geometry for the interacting terms appearing in the Hamiltonian of Eq.~\eqref{eq:tfim2d}, including bonds spanning periodic boundaries. The benchmarking of the algorithm for the energy in the $6\times 6$ two-dimensional torus configuration at the critical field value and for different bond dimensions is shown in Fig.~\ref{fig:summary}(e). We see that the ground-state energy improves even for those small bond dimensions chosen.

In contrast to 1D, where DMRG converges relatively well for the noncritical point as compared with the critical point, for the 2D systems the entanglement complexity governed by the bond dimension persists throughout the field range. Therefore, in addition to the critical point, we analyzed the optimization routines for noncritical values both above and below $h_c$. The optimized DMRG results for the system size of $8\times 8$ with the torus geometry are depicted in Fig.~\ref{fig:energy_vs_BD_2d}. We find that the dependence of the energy on the bond dimension exhibits qualitatively distinct scaling behavior across different values of the field. For the point below the critical one, $h=1.0$ (Fig.~\ref{fig:energy_vs_BD_2d}(a)), the improvement in ground-state energy systematically increases with the bond dimension in the $10^{-4}$ range. At $h_c$, the proposed algorithm performs better for larger bond dimensions, as the accuracy of the energy approximation with CAMPO increases with the bond dimension in the $10^{-2}$ range; see Fig.~\ref{fig:energy_vs_BD_2d}(b). Though the accuracy is on the order of $10^{-1}$, at points above the critical one ($h=5.0$), there is almost constant improvement (for large $\chi$) in energy with the bond dimension, as in Fig.~\ref{fig:energy_vs_BD_2d}(c). 

 \begin{figure*}[htb!]
    \centering
    \includegraphics[width=1.0\linewidth]{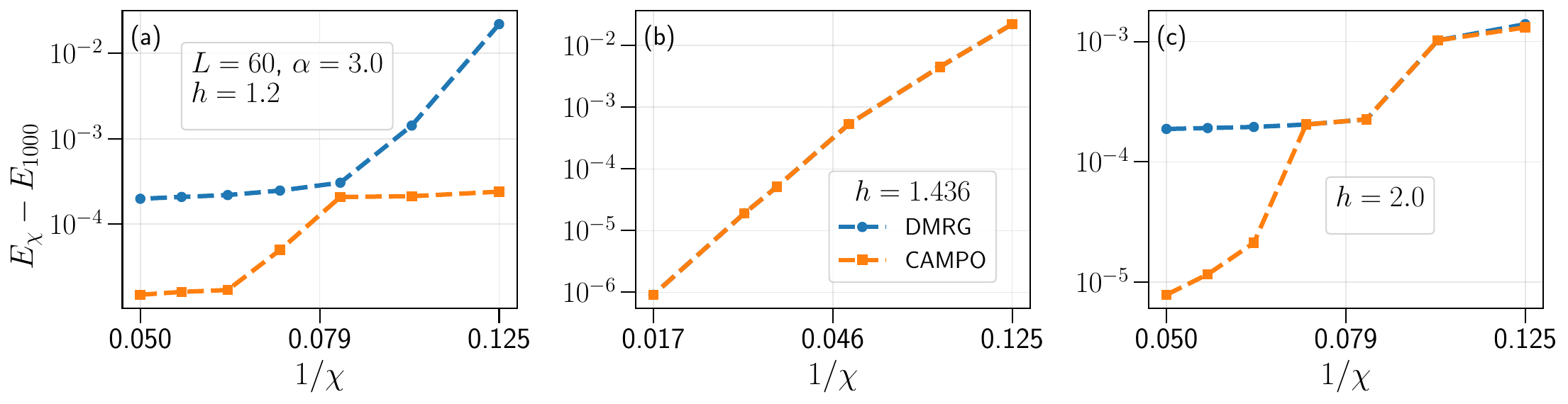}
    \caption{Scaling of the energy difference (relative to the $\chi=1000$ reference value) as a function of $1/\chi$ for standard DMRG and CAMPO, at fixed $L=60$, $\alpha=3$, for three representative values of $h=$ (a) $1.2$, (b) $1.436$($h_c$), and (c) $2.0$.}
    \label{fig:LRTFIM}
\end{figure*}

\textit{1D long-range Ising model}---Next, as an extension of the nearest-neighbor 1D TFIM,  we consider a system with interactions beyond nearest neighbors, where the coupling strength also depends on the pair of sites involved. To this end, we focus on the 1D long-range Ising model with all-to-all spin–spin interactions with power-law decay with the distance between spins
\begin{equation}
    \calH_{\rm LR} = -\sum_{i<j}^L\frac{J}{r_{ij}^{\alpha}}X_i X_j - h\sum_{i=1}^{L}Z_i ,
    \label{eq:lrim1d}
\end{equation}
where $r_{ij} = |i-j|$ denotes the minimum distance between spin sites in the periodic configurations.
The phase diagram of the model described by Eq.~\eqref{eq:lrim1d} in the $\alpha$ vs. $h/J$ plane is governed by the competition between the algebraically decaying ferromagnetic couplings $\alpha$ (which drive $\mathbb{Z}_2$ symmetry breaking) and the transverse field $h$ (which drives quantum fluctuations).
The universality class of the transition is fundamentally dictated by how rapidly the interactions decay across the lattice. The model falls under mean-field criticality for highly extended couplings ($\alpha\leq 1.5$) and under the local Ising universality class for rapidly decaying ones ($\alpha \geq 3.0$) \cite{FisherPRL1972, dutta2001longrange, Vodola_2015, Gonzalez_Lazo_2021, defenu2023longrange}, separated by an intermediate regime characterized by non-local critical behavior and continuously varying exponents. The entanglement structure of this model satisfies an area law for noncritical ground states at sufficiently large $\alpha$~\cite{Kuwahara_2020}, with at most logarithmic violations at criticality in that regime~\cite{vidal2003entanglement, calabrese2004entanglement}; for small $\alpha$, however, entanglement can grow algebraically or even linearly with subsystem size~\cite{PhysRevLett.109.267203}, placing MPS efficiency out of reach there. We test CAMPO at $\alpha=3.0$, the boundary between the intermediate- and short-range regimes, where $h_c\approx1.436$~\cite{PRBKoziol}. In addition to $h_c$, we consider $h=1.2$ and $h=2.0$, below and above the transition, respectively. Figure~\ref{fig:LRTFIM} shows the energy difference for the DMRG and CAMPO results relative to the $\chi=1000$ DMRG energy, versus $1/\chi$. The improvement is $h$-dependent: in the ordered phase CAMPO consistently beats DMRG by one to two orders of magnitude as $\chi$ drops from 20 to 8; in the disordered phase it matches DMRG for $\chi\le14$ but gains about two orders of magnitude at larger $\chi$; at $h_c$ it never surpasses DMRG over the range studied ($\chi=8$--60), requiring larger $\chi$ or a more general gate search beyond the brick-wall structure. Overall, CAMPO gives a clear, at times order-of-magnitude, advantage away from criticality, though in the disordered phase only once $\chi$ is large enough.

\textit{Conclusions}---In this work, we have introduced an algorithm for systematically improving the variational energy obtained from DMRG within a fixed-bond-dimension tensor-network ansatz. The central motivation is that, although DMRG provides highly accurate variational states, the finite bond dimension can leave substantial room for further optimization. Our approach exploits this variational freedom by updating the Hamiltonian through Clifford-unitary transformations while keeping the MPS bond dimension fixed throughout. 
We benchmark the method on spin systems with periodic boundary conditions in both one and two dimensions. For systems whose ground states obey an area law, we observe a striking improvement in the energy compared to the plain DMRG one, with improvement gains of order $\sim 10^{-4}$. For systems with long-range entanglement, where the corresponding one-dimensional MPS representation exhibits volume-law scaling, the improvement is generally more modest, but remains significant at intermediate bond dimensions. 

Beyond energy optimization, this framework provides a useful route to studying Hamiltonian dynamics in the Heisenberg picture while keeping the complexity of the underlying MPS representation fixed. Further, the algorithm can be combined with other approaches such as CAMPS~\cite{PhysRevLett.133.190402} and Pauli propagation~\cite{rudolph2025pauli} to study phenomena like entanglement dynamics, correlations, time evolution, and many others efficiently. At a broader level, one can think about extending our framework to incorporate larger (e.g., 3-body) Clifford augmentation, possibly utilizing stochastic optimization.

\textit{Acknowledgments}---The tensor network calculations are done with \texttt{ITensors}~\cite{itensor, itensor-r0.3}, and the Clifford update is performed using \texttt{STIM}~\cite{Gidney2021stimfaststabilizer} library. 
We thank Z. Bacciconi, M. Frau, and G. Fux for discussions and collaboration on related projects.
This work was partly supported by the EU-Flagship programme Pasquans2 (M.D.), by INFN Iniziativa Specifica Quantum, by the ERC Consolidator grant WaveNets (Grant agreement ID: 101087692) (M.D.), by ONR Global (M.D.), by the PNRR MUR project PE0000023-NQSTI (M.C. and  M.D.) and by the QUASAR-FVG project founded within the ``Piano Sviluppo e Coesione FVG'' (M.C.).


\end{document}